\documentclass[fleqn,usenatbib]{mnras}

\usepackage{newtxtext,newtxmath}
\usepackage{longtable}

\usepackage[T1]{fontenc}

\DeclareRobustCommand{\VAN}[3]{#2}
\let\VANthebibliography\thebibliography
\def\thebibliography{\DeclareRobustCommand{\VAN}[3]{##3}\VANthebibliography}

\usepackage{graphicx}	
\usepackage{amsmath}	
\usepackage{threeparttable}
\usepackage{soul}
\usepackage{xcolor}

\title[Revised CaT Calibration]{Revisiting the near infrared Calcium triplet as metallicity indicator}

\author[M. Navabi et al.]{
M. Navabi,$^{1}$\thanks{E-mail: m.navabi@surrey.ac.uk}
R. Carrera,$^{2}$
N. E. D. No\"el,$^{1}$
C, Gallart,$^{3,4}$
E. Pancino,$^{5,6}$
M. De Leo$^{7,2}$
\\
$^{1}$Department of Physics, University of Surrey, Guildford GU2 7XH, Surrey, UK\\
$^{2}$INAF-Osservatorio di Astrofisica e Scienza dello Spazio, via P. Gobetti 93/3, 40129, Bologna, Italy\\
$^{3}$Instituto de Astrof\'{\i}sica de Canarias, La Laguna E-3200, Tenerife, Spain\\
$^{4}$Departamento de Astrof\'{\i}sica, Universidad de La Laguna, La Laguna E-38205, Tenerife, Spain\\
$^{5}$INAF – Osservatorio Astrofisico di Arcetri, Largo E. Fermi 5, I-50125 Florence, Italy\\
$^{6}$Space Science Data Center – ASI, Via del Politecnico SNC, I-00133 Roma, Italy\\
$^{7}$ Dipartimento di Fisica e Astronomia, Universita degli Studi di Bologna, Via Piero Gobetti 93/2, Bologna, 40129, Italy\\
}

\date{Accepted XXX. Received YYY; in original form ZZZ}

\pubyear{2025}

\begin{document}
\label{firstpage}
\pagerange{\pageref{firstpage}--\pageref{lastpage}}
\maketitle

\begin{abstract}

The near-infrared Calcium\,\textsc{ii} Triplet (CaT), around 850\,nm, is a key metallicity indicator for red giant stars. We present a revised [Fe/H] calibration as a function of CaT line strengths and four luminosity indicators, including the \textit{Gaia} \textit{G}-band, together with the classical $V$, $I$, and $K_s$ bandpasses. For this purpose, we used a sample of 366 red giant stars belonging to 25 globular and open clusters,  complemented by 52 extremely metal-poor field giant stars. 
The CaT line strengths are determined by fitting Gaussian–Lorentzian combination profiles using the \textsc{Python} \textsc{lmfit} package, which utilises the algorithms implemented therein.
The derived calibration is valid for a wide metallicity range, $-4$\,dex$ \lesssim \mathrm{[Fe/H]} \lesssim +0.15$\,dex, and for ages older than $\sim$200\,Myr. In addition, we performed a detailed assessment of how factors such as spectral resolution, spectral quality (expressed through the signal-to-noise ratio), and the algorithms used to constrain the line profiles affect the measured line strengths and the resulting metallicities.

\end{abstract}

\begin{keywords}
stars: abundances — stars: late-type —
\end{keywords}


\unskip
\begin{table}
\setlength{\tabcolsep}{0.9mm}
\renewcommand{\arraystretch}{1.2} 
\caption{Reference values for the stellar clusters in our sample.}
\centering
\label{tab:clusters}
\begin{tabular}{lccccccccc}
\hline
            Cluster     &    [Fe/H]       & Ref& E(B - V ) & (m-M)$_0$ & Ref\\  
\hline            
    NGC 104 (47 Tuc)& -0.76 $\pm$ 0.02 & 1& 0.04 & 13.27 $\pm$ 0.01&   2,3  \\
    NGC 288       & -1.32 $\pm$ 0.02 & 1& 0.03 & 14.76 $\pm$ 0.02  &  2,3\\
    NGC 362       & -1.09 $\pm$ 0.04 & 8 & 0.05 & 14.73 $\pm$ 0.02 &  2,3\\
    NGC 1851      & -1.18 $\pm$ 0.08 & 1& 0.02 & 15.38 $\pm$ 0.02   &  2,3\\
    NGC 1904 (M79)& -1.58 $\pm$ 0.02 & 1& 0.01 & 15.58 $\pm$ 0.01 &  2,3\\
    NGC 2298      & -1.96 $\pm$ 0.04 & 1& 0.14&  14.96 $\pm$ 0.04&   2,3\\
    NGC 3201      & -1.51 $\pm$ 0.02 & 1&0.24 & 13.37 $\pm$0.02 &  2,3\\
    NGC 4590 (M 68)& -2.27 $\pm$ 0.04& 1& 0.05 & 15.08 $\pm$ 0.03  &  2,3\\
    NGC 5927      & -0.29 $\pm$ 0.07 & 1& 0.45 & 14.58 $\pm$ 0.03 &  2,3\\
    NGC 6352      & -0.62 $\pm$ 0.05 & 1& 0.22 & 13.72 $\pm$ 0.03  &  2,3\\
    NGC 6528      & +0.07 $\pm$ 0.08 & 1& 0.54 & 14.47 $\pm$ 0.07 &  2,3\\
    NGC 6681      & -1.62 $\pm$ 0.08 & 1& 0.07 & 14.85 $\pm$ 0.02&  2,3 \\
    NGC 7078 (M 15)& -2.33 $\pm$ 0.02 & 1& 0.10 & 15.15 $\pm$ 0.02 &  2,3\\ 
\hline
    Berkeley 17   & -0.24 $\pm$ 0.04&  5 & 0.06  & 12.38 $\pm$ 0.81   & 4\\
    Berkeley 20   & -0.38 $\pm$ 0.02 &  6 & 0.12 & 13.25 $\pm$ 5.64  & 4 \\
    Berkeley 39   & -0.14 $\pm$ 0.01& 6  & 0.06  & 12.90 $\pm$ 1.11   & 4 \\
    Collinder 110 & -0.10 $\pm$ 0.02&  6 & 0.37  & 11.69 $\pm$ 0.32   & 4 \\ 
    Melote 66     & -0.33 $\pm$ 0.03& 7  & 0.08 & 13.20  $\pm$ 0.91  & 4 \\
    NGC 188       & -0.03 $\pm$ 0.07 &5 & 0.07  & 11.15 $\pm$ 0.19  &  4\\
    NGC 2141      & -0.04 $\pm$ 0.16&  6 & 0.31  & 12.91 $\pm$ 1.28   &  4\\
    NGC 2682 (M67)& +0.04 $\pm$ 0.04& 5 & 0.02  &  09.75  $\pm$ 0.10   & 4\\
    NGC 6705 (M11)& +0.11 $\pm$ 0.07& 5 & 0.39   & 11.72 $\pm$ 0.42   &  4\\
    NGC 6791      & +0.15 $\pm$ 0.14& 5 & 0.22  & 13.13  $\pm$ 0.95  & 4 \\
    NGC 6819      & +0.04 $\pm$ 0.06& 5 & 0.13   & 12.21 $\pm$ 0.28   & 4 \\
    NGC 7789      & +0.00 $\pm$ 0.07& 5 & 0.26 & 11.60   $\pm$ 0.23 & 4 \\
\hline 
\end{tabular}
\begin{tablenotes}
 \item[a] (1) \cite{2009CarrettaC} ; (2) \cite{Harris2010}; (3) \cite{2021Baumgardt}; (4) \cite{2020Cantat}; (5) \cite{2024Carbajo}; (6) \cite{2022A&A...666A.121R}; (7) \cite{2014Carraro}; (8) \cite{2022Vargas}
\end{tablenotes}
\end{table}

\unskip

\section{Introduction}

The near-infrared \ion{Ca}{ii} triplet (CaT) lines, located at 849.8, 854.2, and 866.2\,nm, respectively, are easily distinguishable even in low and medium-resolution near-infrared spectra of late-type giant stars \citep[e.g. ][]{Armandroff1991,2001Cenarro}. Due to their high sensitivity to changes in stellar metal content, these lines have become widely used as metallicity indicators for a range of stellar systems. This includes both old metal-poor systems such as globular clusters and dwarf spheroidal galaxies  \citep[e.g.,][]{Armandroff1988, 1997Rutledge, Battaglia2008, Lucchesi2020, Sakari2022}, as well as metal-rich open clusters or dwarf galaxies like the Magellanic Clouds \citep[e.g.,][]{Costa1998, Tolstoy2001, Cole2004, 2009Leaman,Parisi2010, Parisi2022,Olszewski1991, Carrera2008A, Carrera2008, Carrera2011, Bortoli2022,2012A&A...544A.109C,2015Carrera,2017Carrera}. 

The strength of the CaT lines is quantified by the area between the line profile and the continuum level. The stellar chemical abundance, effective temperature, and surface gravity mainly influence this strength.
The pioneering works by \citet{Armandroff1988, Armandroff1991, Olszewski1991}, focused on red giant stars belonging to Galactic and Large Magellanic Cloud globular clusters, noted that the effective temperature and surface gravity dependences could be removed by assuming a linear relation between the strengths of the CaT lines, and a luminosity indicator such as the absolute magnitude in \textit{V}-band or the difference between the magnitudes of the stars and the horizontal branches also in \textit{V}-band. The strength of the lines was determined by fitting a Gaussian function to their profiles.
In order to extend the use of the CaT as a metallicity indicator to more metal-rich and younger regimes, \citet{Cole2004} proposed to model the line profiles with a combination of a Gaussian and a Lorentzian functions, since they provide a more accurate fit to the core and wings of these lines. This procedure also removes potential blends with other lines, particularly important in the extended wings of more metal-rich objects.
This approach was successfully applied by \citet{Carrera2007}, extending the use
of the CaT lines as a metallicity indicator for metal-rich stars, up to $+0.15$\,dex. 

On the other side, \citet{Starkenburg2010} extended the CaT calibration to extremely metal-poor regimes using synthetic spectra. They noticed that the relationship between the strength of the CaT lines and the luminosity indicators is not linear, and the addition of non-linear terms is needed to properly reproduce the observed trend. Taking this into account, \citet[][hereafter C13]{2013Carrera} complemented the open and globular cluster sample used by \citet{Carrera2007} with 50 extremely metal-poor stars with [Fe/H] $\lid~-2.5$\,dex, obtaining an empirical calibration expanding a wide range of metallicities, $-4 <$[Fe/H]$< +0.15$, and ages, $\geq$0.25\,Gyr using \textit{V}, \textit{I} Johnson-Cousins.
This calibration has been widely used in the literature \citep[e.g. ][]{2014Mauro,2014Koch,2015Ho,2015Carrera,2017Carrera,2015Simon,2020Simon,2015Slater,2017Li,2018Li,2019Longeard}, also for obtaining the metallicity of ultra-faint dwarf galaxies, more recently \citep{2021ApJenkins,2023Cerny,2024Smith,2024Heiger}. 

Our initial aim was to extend the C13's calibration to the \textit{Gaia} \textit{G}-band magnitudes, which, since the first data release \citep{2016Gaia}, have been widely used. In the latest data release, the third one, \textit{Gaia} provided \textit{G}-band magnitudes for 1.7 billion stars, with a precision which reaches the milli-magnitudes \citep[$Gaia$ EDR3,][]{Gaia2021}.
Through this process, we noticed some differences in the line strengths measured on the same spectra by C13 when using the current version of \textsc{IDL}. Interestingly, the results from \textsc{IDL} seem to be quite similar to those generated by the latest \textsc{Python} implementation.
We discuss this in detail in Sect.~\ref{sec:index}. This motivated us to measure the strengths of the CaT lines again and compute a new metallicity calibration. For this, we take advantage of the new values, such as distance, [Fe/H], etc., available in the literature in the last years, mainly, from the different \textit{Gaia} data releases and the complementary spectroscopic surveys (see Sect.~\ref{sec:data})

The structure of this paper is as follows. Section~\ref{sec:data} provides an overview of the observational material used in this study and details updates to the C13 catalogue. Section~\ref{sec:index} describes the methodology for calculating CaT line strengths using \textsc{Python} code and compares them with previous C13 measurements. Section~\ref{sec:calibartion} presents the new calibration of CaT lines as a metallicity indicator. We also validate the derived CaT metallicity against high-resolution reference data and compare our calibration with that of C13. Finally, Section~\ref{sec:summary} summarises the main findings of this paper.

\begin{figure*}
\centering
\includegraphics[width=\textwidth]{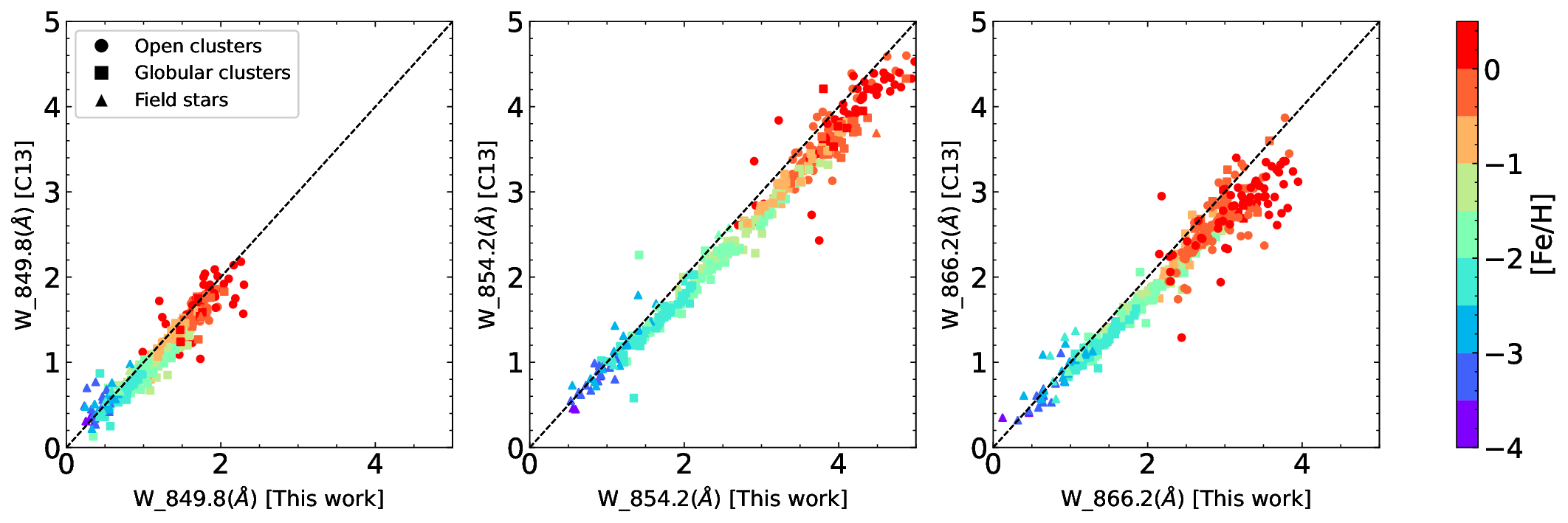}
\caption{Comparison between the C13's strengths derived with the new \textsc{Python} implementation but using the Levenberg–Marquardt algorithm, for each line. Points are colour-coded as a function of the star metallicity, as noted in the right sidebar. Point shapes denote different sources, as labelled in the legend. The dashed lines of equal strength are added for reference in black.}
    \label{figs_metal:eqw_comparing_ind}
\end{figure*}

\section{Observational material}
\label{sec:data}

We use the same data sample as C13, fully described there and by \citet{Carrera2007}.  After applying similar quality constraints as C13, the final dataset consists of 123 red giant stars belonging to twelve open clusters, 243 red giant stars from thirteen globular clusters (see table~\ref{tab:clusters}), and 50 very metal-poor field stars with metallicities [Fe/H] $<$-2.5\,dex. Overall, the sample covers a metallicity range from $-4$ to $+0.15$\,dex. 
C13 employed this sample to establish a relationship between $\Sigma Ca$, defined as the sum of the strengths of the three CaT lines, and metallicity for four luminosity indicators: \textit{V}, \textit{I}, \textit{K$_s$}, and $V - V_{HB}$. In this work, we used as luminosity indicators the same \textit{V}, \textit{I}, and \textit{K$_s$} magnitudes, thoroughly outlined by C13, together with the \textit{Gaia} \textit{G}-band magnitudes \citep{Gaia2021}. We do not attempt to derive a new calibration for $V - V_{HB}$ due to the challenges in defining the horizontal branch position in field stars, poorly populated clusters, or galaxies with extended star formation histories. 

To obtain the \textit{G} magnitude for all stars in our sample, we cross-matched it with \textit{Gaia} EDR3, using as a selection criterion a separation threshold $\leq$ 1\,arcsec by the query in Gaia Archive \footnote{\url{https://gea.esac.esa.int/archive/}}. Additionally, we retrieved the \textit{Gaia} IDs by matching our catalogue with the \textsc{SIMBAD} database \citep[][]{simbad}, based on the \textsc{SIMBAD} identifiers. For the majority of the stars both procedures converge in identifying the same pairs. For ten stars with discrepant results, we manually verified their positions using the \textsc{Aladin} sky atlas \footnote{\url{https://aladin.cds.unistra.fr/}} to identify the correct pairs. Finally, neither procedure can get a match for eighteen globular cluster stars. Their \textit{G} magnitudes were also obtained through a manual search using \textsc{Aladin}.

The absolute magnitudes are calculated as: $M_{i} = m_{i} - A_i - \mu $ with $i=V, I,K_{s}, G$, where $m_{i}$ and  $A_i$ are the apparent magnitude and extinction for \textit{i}-bandpass, respectively, and $\mu$ is the distance modulus.
For open clusters, we adopt the distance moduli and reddening values, $E(B-V)$, provided by \citet{2020Cantat}, which are based on \textit{Gaia} DR2 \citep{Evans2018}.
For the globular clusters, distances derived directly from \textit{Gaia} parallaxes should be affected by large uncertainties due to the large distances for most of these systems, but also by the crowding in their central regions. For this reason, the distance moduli used are derived from the distances determined by \citet{2021Baumgardt} obtained by averaging results provided by different methods. The reddening values are obtained from the Harris globular cluster database\footnote{\url{https://heasarc.gsfc.nasa.gov/W3Browse/star-catalog/globclust.html}} \citep[][]{harris_gc_database2010}.
For field stars, we use the same reddening values adopted in C13. The distance moduli, however, are computed from the latest \textit{Gaia} EDR3 parallaxes, corrected for parallax systematics following the procedure outlined in \citet{Lindegren2021}.

The extinctions, $A_i$, are derived from reddening as: $A_i = E(B-V) * R_V* \kappa_i$, assuming $R_V$=3.1\,mag and $\kappa_i$=1.0, 0.470, and 0.114\,mag~for \textit{V}, \textit{I}, and $k_s$-bandpasses, respectively \citep{1989ApJ...345..245C}.
Owing to the fact that to \textit{Gaia} \textsc{G} magnitudes are derived from a very broad filter, the extinction coefficient depends not only on the extinction itself, but also on the spectral energy distribution of the source \citep{2016Gordon,2018Danielski}. In order to take all of these into account, we use \textsc{Dustapprox} \citep{Fouesneau_dustapprox_2022} a tool developed by the \textit{Gaia} team to derive extinction in \textit{G}-band from $A_V$ and effective temperature of the source. The former is derived from the $(G_{BP}-G_{RP})$ colour provided by \textit{Gaia}, yielding residuals of about 5\% in temperature for stars hotter than 4,500\,K \citep{2010Jordi}. Although most of the stars in our sample have more accurate effective temperatures derived from other methods such as high-resolution spectroscopy, we rely on \textit{Gaia} data for consistency. On one side, the relationships used by \textsc{Dustapprox} have been obtained using the same procedure. On the other, our final goal is to apply the derived calibration to determine metallicities for sources without alternative determinations of the effective temperatures, for which we will rely in most cases only on \textit{Gaia} results.

The reference metallicity values for globular clusters and field stars are the same as those used by C13 since they do not have more recent chemical abundance determinations in the literature. NGC~362 is the only exception, with a very recent determination by \citet{2022Vargas}. For open clusters, we updated the values with the recent high-resolution determinations provided by \citet{2022A&A...666A.121R} and \citet{2024Carbajo} from GES and OCCASO surveys, respectively. The exception is Melotte~66 for which we use the determination provided by \citet{2014Carraro} since it has not been observed by any of the other two surveys. In comparison with C13, we highlight the significant change of metallicity for NGC~6791, from [Fe/H]=+0.47\,dex \citep{2007A&A...473..129C} to +0.15\,dex \citep{2024Carbajo} in agreement with other recent determinations for this cluster \citep[e.g.][]{2022AJ....164...85M}. All the used reference values are listed in Tables~\ref{tab:clusters} and \ref{tab:ref_field_stars} for clusters and field stars, respectively.

\section{The C\lowercase{a}T index}
\label{sec:index}

The strength of the CaT lines has been obtained following a similar procedure as C13, but with some modifications detailed below. Briefly, the strength of every line is quantified as the area between the line profile inside a bandpass covering the feature and the continuum level evaluated in several bandpasses between the three CaT lines. For this purpose, we use the bandpasses defined by \citet{2001Cenarro}. The profile of each CaT line is fitted to a combination of a Gaussian and a Lorentzian function as proposed by \citet{Cole2004}. We refer the reader to C13 and \citet{Carrera2007} for a detailed discussion of these choices. 

The strength of each CaT line is quantified by fitting its profile with a combination of a Gaussian and a Lorentzian function, following previous work, using a least-squares fit. C13 and \citet{Carrera2007} used the \textsc{IDL} implementation of \textsc{lmfit} package, using the Levenberg-Marquardt algorithm. During the preparation of this work, we tried to use the same implementation but we found significant discrepancies: the strengths derived from C13's original code and algorithm produced different results when run on the latest version of \textsc{IDL}. To investigate the source of this issue, we developed a new, independent code in \textsc{Python} using its implementation of the \textsc{lmfit} package \citep{lmfit,newville_2025_lmfit}. The line strengths obtained with the new \textsc{Python} code are consistent with those derived from the updated \textsc{IDL} version, but both disagree systematically with the original C13 measurements. The new strengths are generally larger than the C13 values, and this deviation is more pronounced for the strongest lines, accompanied by a larger scatter, as shown in Fig.~\ref{figs_metal:eqw_comparing_ind}. This systematic difference is particularly significant for the reddest line at 866.2\,nm. We attribute this sensitivity to the presence of a strong \ion{Fe}{ii} line at approximately 867.5\,nm. For metallicities above about -0.25\,dex, this contaminant line strengthens significantly, complicating the fitting of the line wings and increasing the scatter in the derived strength.

Given its widespread use and public availability, we proceeded with the analysis using the newly developed \textsc{Python} code. Since the \textsc{Python} implementation of \textsc{lmfit} readily allows for testing various optimisation algorithms, we explored several alternatives (see Sect.~\ref{sect:robustness} for details). We found that the \textsc{Nelder-Mead} algorithm yields better fits to our spectra than \textsc{Levenberg–Marquardt}, especially in the wings of metal-rich objects where line blending is a concern.  Furthermore, the flexible \textsc{lmfit} \textsc{Python} implementation allowed us to incorporate Markov Chain Monte Carlo (MCMC) routines through the \textsc{emcee} package, to derive a more robust and realistic quantification of the uncertainties in the fitted line profiles. Finally, the line profile area is calculated using the \textsc{Newton-Cotes} integration rule from the \textsc{SciPy} package \citep[][]{2020SciPy-NMeth}. This code is publicly available on \textsc{Gitlab}\footnote{\url{https://github.com/carrerajimenez/cat_pipeline/}}.

\begin{figure}  
\includegraphics[width=\columnwidth]{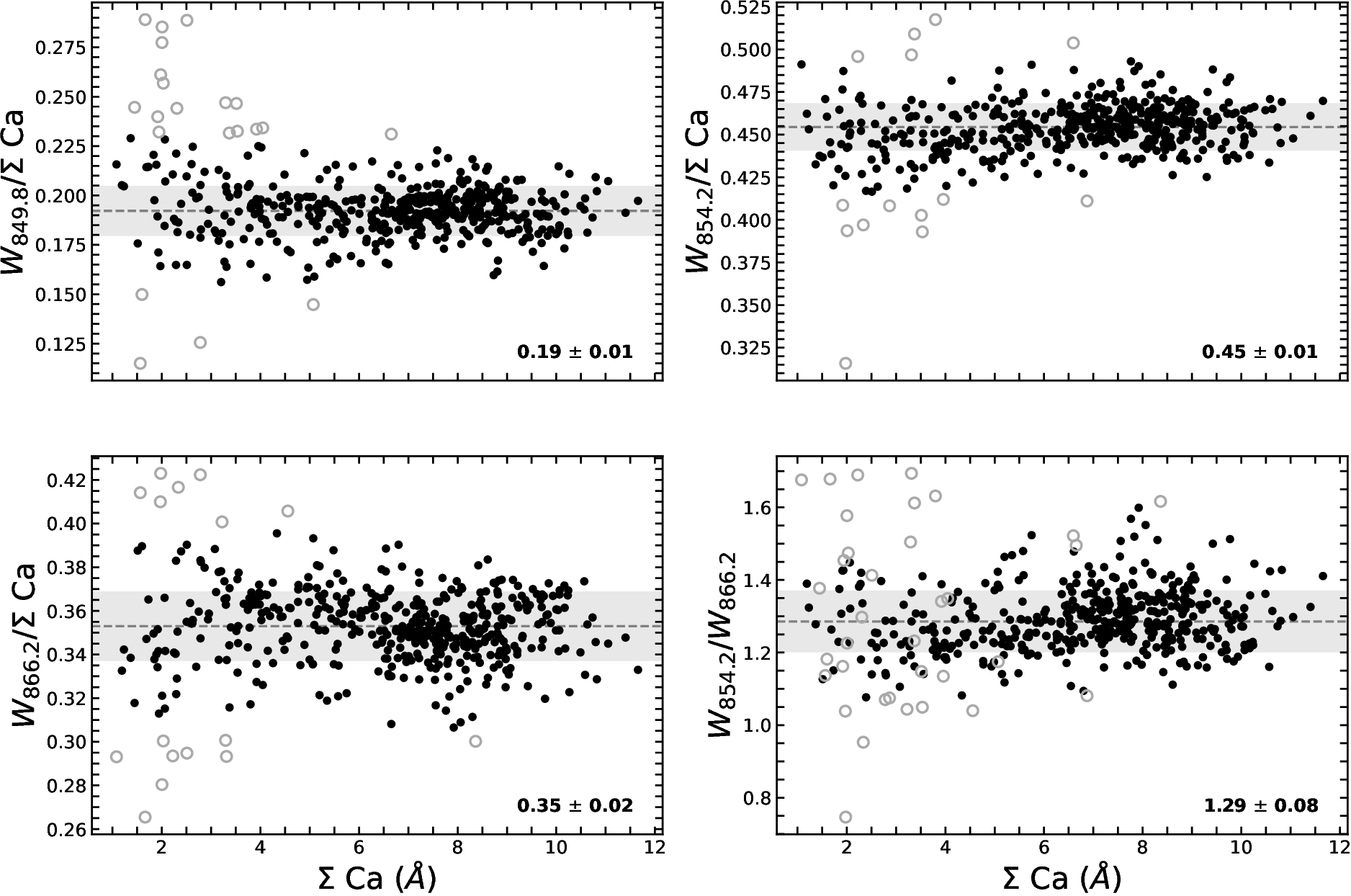}
       \caption{Contribution of the strengths of 849.8\,nm (top-left), 854.2\,nm (top-right), and 866.2\,nm (bottom-left) to the global CaT index, $\Sigma$~Ca. Bottom-right panel show the behaviour of the $W_{854.2}$/$W_{866.2}$ ratio. Median (dashed lines) and standard deviation (shadow regions) shown in the bottom-right corner of each panel have been computed applying a three-sigma clipping (open circles show rejected points).}
\label{fig:line_contribution}
\end{figure}

Finally,  the CaT index, $\Sigma Ca$, is simply derived as the sum of the strengths of the three lines: $\Sigma Ca=W_{849.8}+W_{854.2} +W_{866.2}$. 
We point the reader to C13 and \citet{Carrera2007} for the discussion of this selection in comparison with other approaches used in literature. Other prescriptions have been used in the literature such as excluding the weakest line at 849.8\,nm \citep[e.g.][]{1993ApJ...418..208S,Battaglia2008}, or assigning different weights to each line \citep[e.g.][]{1997Rutledge}. C13 studied the contribution of each line to the CaT index. We confirm in the present study that the relationships they derived remain valid when the new procedure is applied. The weakest line at 849.8\,nm contributes only $19\pm1$\% to $\Sigma Ca$, while the other lines at 854.2 and 866.2\,nm contribute with $45\pm1$\%, $35\pm2$\%, respectively (see Fig.~\ref{fig:line_contribution}). The sum of the two strongest lines is responsible of $81\pm1$\% of the total index.  Furthermore, the ratio between the strength of the two strongest CaT lines, $W_{854.2}/W_{866.2}$, is investigated finding a value of $1.29\pm0.08$, obtained from all the stars in our sample, independently of the absolute magnitude. This value is in good agreement with the results obtained by \citet{Starkenburg2010} from synthetic spectra, which incorporated non-linear thermodynamic equilibrium effects, and \citet{Husser2020}, using the same bandpasses as here but a Voigt profile. However, our data exhibit subtle indications that the proportional contribution of each line to the global $\Sigma Ca$ index is not strictly constant across the entire metallicity range, particularly for stars in the extremely metal-poor regime. This slight deviation is difficult to confirm definitively because these lines are intrinsically very weak for metal-poor objects, leading to higher measurement uncertainties close to the noise limit.

\subsection{Robustness and Sensitivity Analysis}\label{sect:robustness}

\begin{figure*}  
\includegraphics[width=0.9\textwidth]{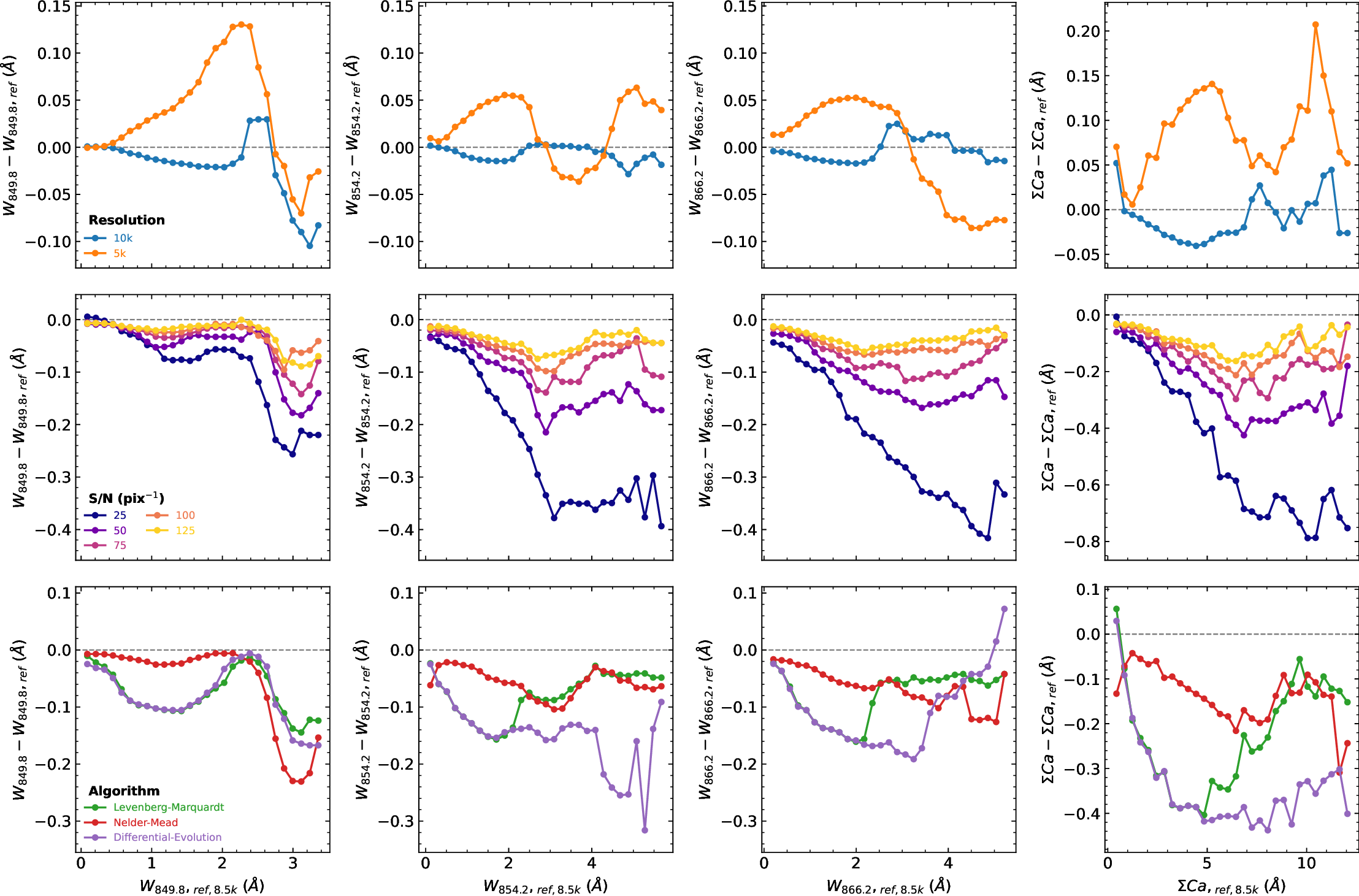}
       \caption{Comparison of the strengths determined for each CaT line (first three left columns) and the global $\Sigma Ca$ index (right) columns determined from synthetic spectra (see text for details) of different resolutions (top row), signal-to-noise ratios (S/N, middle row) and, and using different fitting algorithms (bottom row). As a reference, we use the values determined from synthetic spectra with a spectral resolution of 8,500 using the same methodology, Nelder-Mead plus \textsc{emcee} package, used in the observed spectra.}
    \label{fig:robustness}
\end{figure*}

There are several extrinsic aspects which can affect the accuracy of the determination of the CaT strengths. They are related to the nature and quality of the used spectra, such as spectral resolution or signal-to-noise ratio (S/N), and to the methodology itself, such as the assumed line profile, the continuum determination, the bandpasses used, or the fitting algorithm used. The literature has extensively discussed the impact of different bandpasses and profiles on line strength determination \citep[e.g.,][]{Cole2004,2013Carrera,Vasquez2015}. The Gaussian profile, widely used in the past \citep[e.g.,][]{Armandroff1991,Battaglia2008} works reasonably for the metallicity range covered by the Galactic halo globular clusters. However, more complex profiles, like the Gaussian-Lorentzian combination used here, are necessary for metal-rich stars (e.g., open clusters) or a high spectral resolution ($\lambda / \delta \lambda \ approx$10,000).

To investigate the impact of other factors on the derived strengths, we created a trial sample of synthetic spectra from the homogeneous collection published by \citet{allendeprieto2018}. We refer the reader to this paper for details. We used as reference the \textit{nsc1} library with a spectral resolution of $ \lambda / \delta \lambda$10,000 and three free parameters: effective temperature, $T_{\rm eff}$; surface gravity, $\log g$, and metallicity, [Fe/H]. For our sample we selected spectra with 3,750$\leq T_{\rm eff}/[K] \leq$5,000 with a step of 250\,K; 0.5 $\leq log g \leq $2.0 with a step of 0.2\,dex; and -5.0 $\leq [Fe/H] \leq$0.5\,dex with a step of 0.5\,dex below [Fe/H]=-1.0\,dex and 0.25\,dex above it. We are aware that several of the combinations of these parameters do not match the properties of red giant stars; however, we keep them as we would like to check the behaviour of our methodology in extreme cases.

In the past, several authors have applied the CaT calibration to different datasets obtained with different instrumental configurations and therefore spectral resolutions. Therefore, we first checked the impact of this on the derived strengths. For this purpose, we smoothed our trial sample—originally computed at a spectral resolution of 10,000—to resolutions of 8,500 (matching the observational spectra used as reference) and 5,000. The obtained results are shown in the top row of Fig.~\ref{fig:robustness}. In all the cases, the strengths have been determined using the \textsc{Nelder-Mead} algorithm and refining the results with \textsc{emcee}, as in the case of the observed stars. 
On average, there is no significant difference between the values measured at the two highest resolutions (blue line), even for the strongest lines in the most metal-rich objects. The total $\Sigma Ca$ index differs by less than $\pm$0.05\,\AA, making the impact on the final metallicities negligible. At lower resolution, the differences can exceed 0.15\,\AA, but this only results in the derived metallicity being underestimated by roughly 0.05\,dex.

The quality of the spectra, quantified as the S/N ratio, should have a significant impact on the derived strength because this not only complicates the line profile fitting, but also the continuum determination. In order to investigate its impact, we have added noise to them to reach S/N of 25, 50, 75, 100 and 125\,pix$^{-1}$, respectively. For simplicity, we restricted this analysis only to the 8,500 resolution. The obtained comparisons are shown in the middle row of Fig.~\ref{fig:robustness}. 
A lower S/N ratio leads to larger deviations in the measured line strengths compared to the reference values obtained from the noiseless 
R = 8500 spectra.
The derived strength can be underestimated up to 0.8\,\AA~for the S/N$\sim$25\,pix$^{-1}$, which yields an underestimation of the metallicity of about 0.25\,dex. This differences reduce to 0.4\,\AA~for a S/N$\sim$50\,pix$^{-1}$ and to 0.2\,\AA~for S/N ratio higher than 75\,pix$^{-1}$. They imply differences in the final metallicities of 0.1 and 0.05\,dex, respectively. These larger differences are found for metallicities higher than [Fe/H]$\sim$-0.5\,dex. In the range of globular clusters, S/N plays a minor role; even at the lowest S/N, the maximum deviation is $\sim$0.4\,dex, implying a metallicity underestimation of $\sim$0.15\,dex.
Interestingly, for extremely metal-poor objects where the lines are weakest, the quality of the spectra has a stronger impact on measuring the strength of the reddest line at 886.2\,nm than on the other Ca II triplet lines.

Finally, taking advantage of the flexibility of the \textsc{Python} implementation of the \textsc{lmfit} package, we have investigated the differences in the line strengths using three different algorithms: the classical, gradient-based, \textsc{Levenberg-Marquardt}, the widely used, direct search, \textsc{Nelder-Mead} one, and a genetic \textsc{differential evolution} one. In this case, we determined the strength directly with these algorithms without applying the MCMC analysis with \textsc{emcee}, and in the referenced ones. The obtained results are shown in the bottom row of Fig.~\ref{fig:robustness}. In general, the \textsc{Nelder-Mead} algorithm works reasonably well in the whole range of the strength of the three CaT lines. For this reason, we begin our profile fit with this. The \textsc{Levenberg-Marquardt} serves well for the strongest lines, but it works worse for weaker lines. The genetic algorithm does not appear to accurately reproduce the line profiles using the Gaussian–Lorentzian combination.
In fact, the measured line strengths can be underestimated by up to 0.4\,\AA, which again corresponds to an underestimation in metallicity of roughly 0.15\,dex. Nonetheless, these algorithm-dependent differences are still too small to explain the larger discrepancies with the earlier \textsc{IDL}-based results, as shown in
Fig.~\ref{figs_metal:eqw_comparing_ind}.

\begin{figure*}
\includegraphics[width=0.9\textwidth]{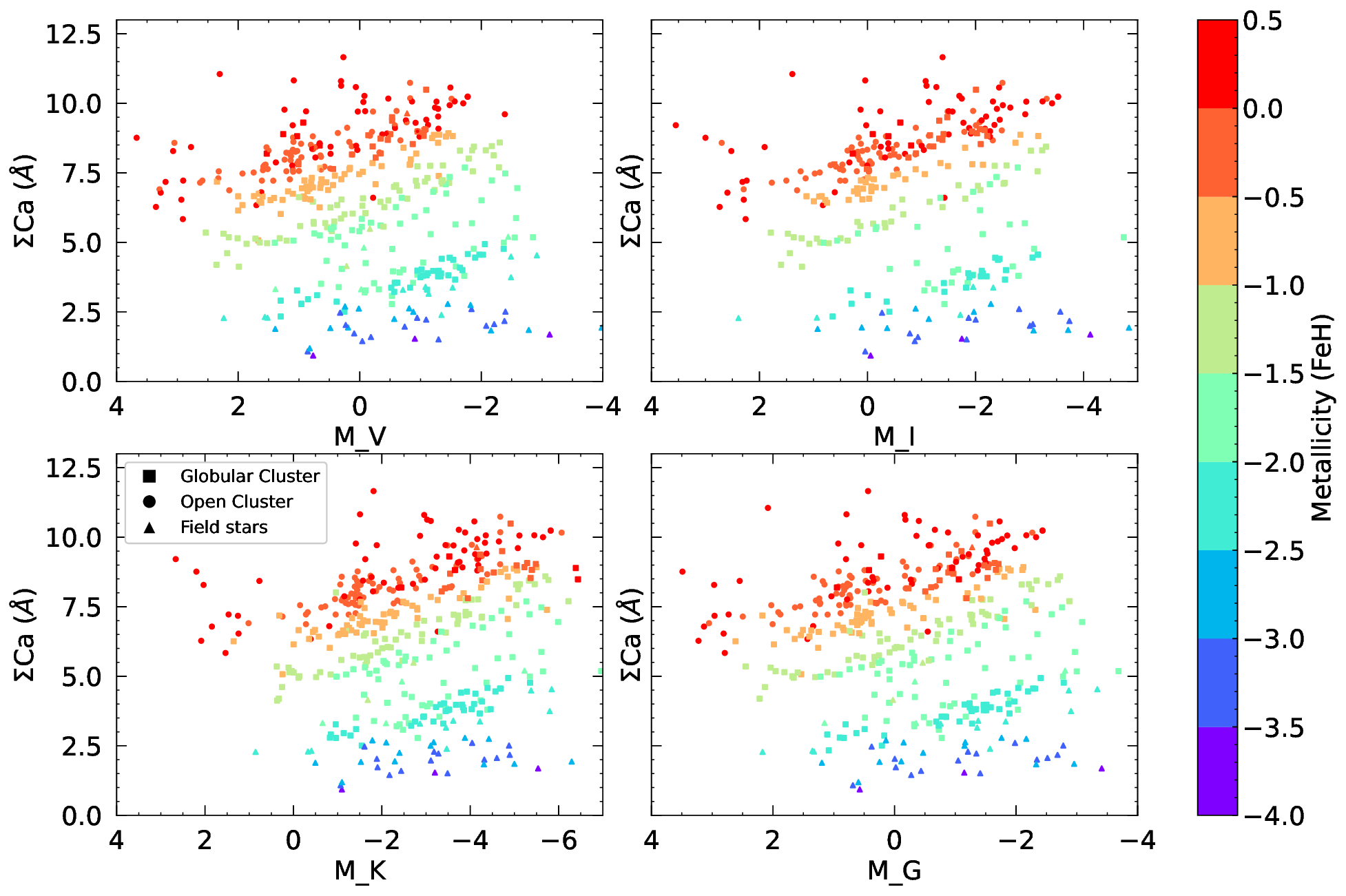}
    \caption{The relationship of the CaT index ($\Sigma Ca$) with the luminosity indicators $M_V$, $M_I$, $M_{K_s}$ and $M_G$, respectively. The colour bar represents the metallicity range for each star. Different point shapes represent three categories of different tracers, as indicated in the legend.}
    \label{figs_metal:samsca_mv_feple}
\end{figure*}

\section{A revised C\lowercase{a}T metallicity calibration}
\label{sec:calibartion}

Figure~\ref{figs_metal:samsca_mv_feple} shows the run of $\Sigma Ca$ as a function of the four luminosity indicators used: $M_I$, $M_{K_s}$, $M_V$, and $M_G$, respectively.
The points are coloured as a function of their metallicity, as indicated in the right-hand sidebar.  
It is clear that the sequences formed by objects with similar metallicities are not linear, contrary to the assumption made in the pioneering studies in this field. \citep[e.g.][]{Armandroff1988,Armandroff1991}.

Different recipes have been used in the literature to account for this non-linearity \citep[e.g.][]{Husser2020}. In this case, following \citet{Starkenburg2010} these sequences are parametrised as:

 \begin{equation*}
    [Fe/H]_{i} = a + b \times M_{i} +c \times \Sigma Ca +d \times \Sigma Ca ^{-1.5} + e \times M_{i} \Sigma Ca
    \label{eq:metal}
\end{equation*}
where $M_{i}$ refers to absolute magnitude in each band analysed: \textit{V}, \textit{I}, $K_s$, and \textit{G}, respectively. This includes the $\Sigma Ca ^{-1.5}$ to account for the extremely metal-poor regime and the cross term for the non-linear trends.

\unskip

\begin{table}
\setlength{\tabcolsep}{1.5mm}
\renewcommand{\arraystretch}{1.2} 
    \caption{Best-fitting parameters and the total number of stars used for each band.}
    \centering
    \begin{tabular}{lcccc}
    \hline
    coefficient  & V &I & K & G  \\
		\hline
         a&-3.10 $\pm$ 0.05& -3.11 $\pm$ 0.07 &-2.94 $\pm$ 0.08 & -3.14 $\pm$ 0.02 \\
         b& 0.09 $\pm$ 0.02& 0.08 $\pm$ 0.02 & 0.11 $\pm$ 0.02 &  0.08 $\pm$ 0.01\\
         c& 0.33 $\pm$ 0.01&0.37 $\pm$ 0.01  &  0.37 $\pm$ 0.01 & 0.35 $\pm$ 0.01\\
         d&-1.01 $\pm$ 0.13&-0.96$\pm$ 0.14 & -0.93 $\pm$ 0.13  & -0.94 $\pm$ 0.12 \\
         e&0.02 $\pm$ 0.01& 0.02 $\pm$ 0.01 & 0.01 $\pm$ 0.01 & 0.02 $\pm$ 0.02\\
         Number&416& 300 &414& 412\\
 		\hline
    \end{tabular}
    \label{tab:coefficient}
\end{table}     

\unskip

The five parameters \textit{a},  \textit{b},  \textit{c},  \textit{d}, and  \textit{e} have been computed following a procedure similar to that performed in the line profile fitting. We used the \textsc{Python} \textsc{lmfit} package with a \textsc{Nelder-Mead} algorithm and an MCMC procedure to derive these coefficients and their uncertainties. Derived coefficients together with the number of stars used for each luminosity indicator are listed in Table~\ref{tab:coefficient}.

\begin{figure*}  
   \includegraphics[width=\textwidth]{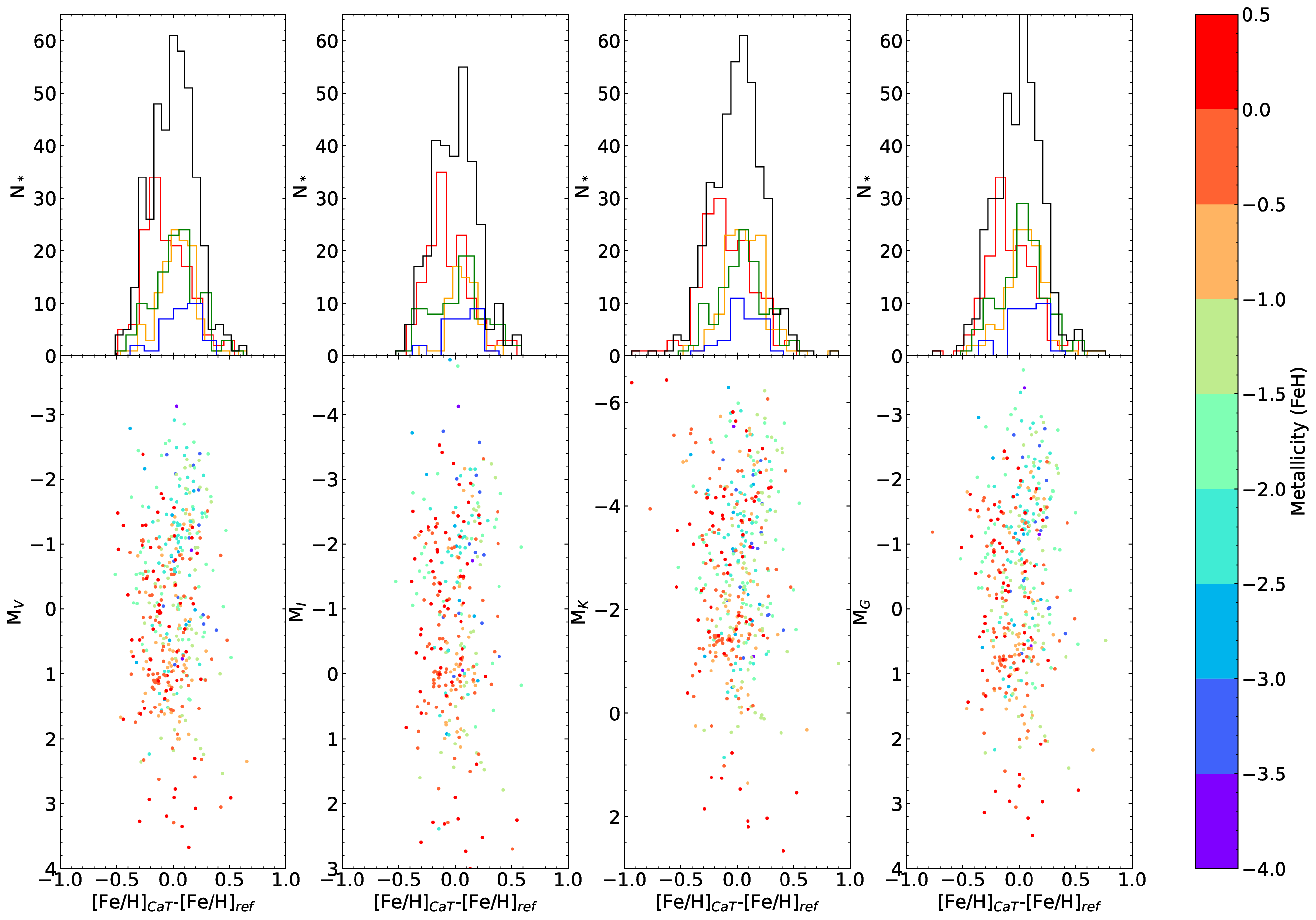}
       \caption{Differences between the reference metallicities and the values derived from the new calibrations obtained here for the four luminosity indicators used (bottom panels), colour coded as a function of metallicity. The top panels show the global distributions of these residuals (black) and in different metallicity ranges (different colours). The statistic of these distributions are summarised in Table~\ref{tab:histogram_res}.}
       \label{fig_metal:histogram_res}
\end{figure*}

To check the accuracy of the derived calibration, we compute the residuals of each star by comparing the reference values with the metallicities obtained using these calibrations for the four luminosity indicators, as shown in the bottom panels of Fig.~\ref{fig_metal:histogram_res}.
Each star has been colour-coded as a function of its metallicity. The top panels show the global distribution of these residuals in black, which are centred at 0.0\,dex, with standard deviations around 0.2\,dex. In principle, this is consistent with the expected uncertainties for this method. However, bottom panels of Fig.~\ref{fig_metal:histogram_res} suggest a trend in the residuals as a function of metallicity, which is quantified by the histograms for different metallicity ranges shown in top panels, and which statistics are listed in Table~\ref{tab:histogram_res}. While the most metal-poor objects show a median difference of about 0.03-0.05\,dex, the most metal-rich ones do it between -0.09 and -0.12\,dex.

An extensive investigation into the source of this trend has been undertaken. Firstly, we restrict the sample to include only stars with absolute magnitudes brighter than zero or filter out a specific group of targets located at the bottom of the red giants in the top-left corner of Fig.~\ref{figs_metal:samsca_mv_feple}. However, the trend persists in all the cases. Furthermore, we explore the addition of second-order and cross terms to the model, such as $M_i^2$, $\Sigma \text{Ca}^2$ or $M_i^2 \times \Sigma \text{Ca}^2$. None of these additions removed the observed trend, as the calculated coefficients are consistently negligible ($\sim 10^{-5}$) compared to the main relation terms. Finally, removing terms already present in the relationship—specifically $\Sigma \text{Ca}^{-1.5}$ or $\Sigma \text{Ca} \times M_i$, resulted in a marked increase in the overall scatter and residuals, confirming the necessity of the current functional form despite the observed systematic behaviour.

\begin{figure*}
\begin{center}
    \includegraphics[width=\linewidth]{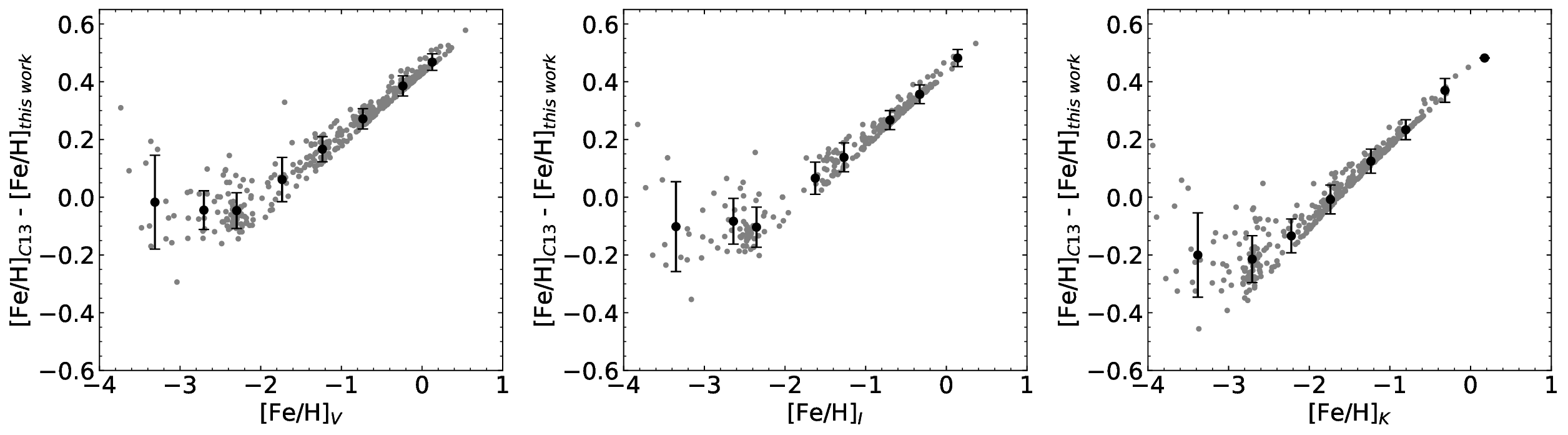}
    \caption{Comparison of metallicities derived from the C13 calibration and the current calibration, using identical absolute magnitudes and line strengths determined in this study.}
    \label{fig_metal:diff_fe}
\end{center}
\end{figure*}

As previously mentioned, the motivation for deriving a new calibration stems from discrepancies observed between strengths derived using the same code but different \textsc{IDL} implementations. Several authors have employed the C13 calibration to derive metallicities in recent years \citep[e.g.][]{giribaldi2023,simon2024}, using strengths determined through the current \textsc{IDL} or \textsc{Python} implementations. To assess the impact of applying the older C13 calibration with these strengths, we compared the metallicities derived from the new calibration with those derived using C13, applying the same magnitudes and CaT strengths from this work. 
Figure~\ref{fig_metal:diff_fe} and Table~\ref{tab:diff_fe} further demonstrate the global impact of this update. The results confirm that both calibrations yield consistent metallicities for [Fe/H]\,$\lesssim$\,$-$1.5\,dex, whereas in the metal-rich regime the older C13 calibration systematically produces higher metallicities by up to $\sim$0.5\,dex. This is in part due to
revised line-strength measurements, but probably mainly to the difference in the reference [Fe/H] values used for NGC~6791, which was +0.47\,dex in C13 but +0.15\,dex here. This updated reference value brings the system into better agreement with the other cluster and ensures that, even when stars from this cluster are removed, the derived coefficients remain stable within the uncertainties.

\unskip
\begin{table}
\setlength{\tabcolsep}{1.5mm}
\renewcommand{\arraystretch}{1.2} 
    \caption{{Median and standard deviation of the residuals' histograms shown in the top panel of Fig.~\ref{fig_metal:histogram_res}.}}
    \centering
    \begin{tabular}{lcccc}
    \hline
    Range  & V &I & K & G  \\
		\hline
         0.5 to -0.5
         &-0.10 $\pm$ 0.19& -0.09 $\pm$ 0.18 & -0.12 $\pm$ 0.22 & -0.12 $\pm$ 0.18 \\
         
         0.5 to -1.5& 0.04 $\pm$ 0.15& 0.07 $\pm$ 0.13 & 0.06 $\pm$ 0.16 &  0.05 $\pm$ 0.15\\
         
         -1.5 to -2.5& 0.04 $\pm$ 0.20&0.05 $\pm$ 0.23  &  0.05 $\pm$ 0.21& 0.04 $\pm$ 0.20\\
         
         > -2.5&0.04$\pm$ 0.17&0.03 $\pm$ 0.17 & 0.05 $\pm$ 0.17  & 0.05 $\pm$ 0.17 \\

         Total& 0.00 $\pm$ 0.19&-0.00$\pm$ 0.20 & 0.00 $\pm$ 0.21  & 0.00 $\pm$ 0.20 \\
 		\hline
    \end{tabular}
    \label{tab:histogram_res}
\end{table}     


\unskip



\unskip

\begin{table}
\setlength{\tabcolsep}{2mm}
\renewcommand{\arraystretch}{1.2} 
    \centering
    \caption{Median and standard deviation of metallicity differences ($\Delta$[Fe/H]) 
    by [Fe/H] Bin. Metallicity differences are calculated between the C13 and present calibrations, using the same absolute magnitudes and line strengths derived in this work.}
 \begin{tabular}{lccc}
        \hline
        [Fe/H] range & $\Delta$[Fe/H]$_{V}$(dex) & $\Delta$[Fe/H]$_{I}$ & $\Delta$[Fe/H]$_{K}$ \\
        \hline
        $\ge 0.0$ & 0.47 $\pm$ 0.03 & 0.48 $\pm$ 0.03 & 0.48 $\pm$ 0.01 \\
        0.0 to $-$0.5 & 0.39 $\pm$ 0.03 & 0.36 $\pm$ 0.03 & 0.37 $\pm$ 0.04 \\
        $-$0.5 to $-$1.0 & 0.27 $\pm$ 0.04 & 0.27 $\pm$ 0.03 & 0.23 $\pm$ 0.03 \\
        $-$1.0 to $-$1.5 & 0.17 $\pm$ 0.04 & 0.14 $\pm$ 0.05 & 0.12 $\pm$ 0.04 \\
        $-$1.5 to $-$2.0 & 0.06 $\pm$ 0.08 & 0.07 $\pm$ 0.06 & $-$0.01 $\pm$ 0.05 \\
        $-$2.0 to $-$2.5 & $-$0.05 $\pm$ 0.06 & $-$0.10 $\pm$ 0.07 & $-$0.131 $\pm$ 0.06 \\
        $-$2.5 to $-$3.0 & $-$0.04 $\pm$ 0.07 & $-$0.08 $\pm$ 0.08 & $-$0.21 $\pm$ 0.08 \\
        $\le -3.0$ & $-$0.02 $\pm$ 0.16 & $-$0.10 $\pm$ 0.16 & $-$0.20 $\pm$ 0.15 \\
        \hline
    \end{tabular}
    \label{tab:diff_fe}
\end{table}

\unskip

\section{SUMMARY}
\label{sec:summary}

In this work, we have revisited the calibration of the CaT as a metallicity indicator, across three luminosity indicators ($M_{V}, M_{I}, M_{K_{s}}$), incorporating the recent widely used \textit{Gaia} $G$-band magnitudes. We have used a sample of 366 red giant stars belonging to 25 clusters, both open and globular, and 52 extremely metal-poor field stars. The revised calibration is applicable across a wide metallicity range, from -4 to +0.15\,dex, for ages older than $\sim$200\,Myr. The residuals of this calibration are within 0.2\,dex, independently of the luminosity indicator used.

The strengths of the CaT lines have been measured with a newly developed \textsc{Python}-based code employing the \textsc{lmfit} and \textsc{SciPy} packages. This code has been made publicly available to the community. We have investigated the contribution of each line to the global $\Sigma Ca$ index, defined as the sum of the strengths of the three lines. Moreover, we have studied the robustness of the line strengths determinations as a function of different factors, such as the spectral resolution and quality, quantified from the signal-to-noise ratio, of the spectra or to the algorithms used to perform the line profile characterisation. Among the various \textsc{lmfit} minimisation methods evaluated, the \textsc{Nelder–Mead} algorithm consistently delivered the best fits across the full range of strengths for all three CaT lines. Also, we conclude that spectral quality has a particularly strong impact on determining the strength of the reddest line (at 886.2\,nm) than on the other spectral lines when observing extremely metal-poor objects.

Finally, we explored the implications of applying the previous C13 calibration alongside the latest determination of the CaT line strengths. Overall, both the new and C13 calibrations yield similar metallicity values for objects with [Fe/H]$\lesssim$ -1.5\,dex, within the associated uncertainties. However, in more metal-rich regimes, the C13 calibration produces higher metallicity values, with a difference of up to $\sim$0.5\,dex compared to the new calibration. This discrepancy arises not only from differences in the CaT line strength measurements, due to the new algorithm implementations, but also from the contrasting [Fe/H] reference value used for NGC~6791, the most metal-rich cluster in our sample.

\section*{Acknowledgements}
We acknowledge the referee, Prof. Gerry Gilmore, for his helpful comments and thoughtful revision of the manuscript, which have contributed to improving the quality and reliability of our results. This work is based on observations obtained from the WHT and the INT, both at the Roque de los Muchachos Observatory (La Palma, Spain); the 4 m telescope at CTIO (La Serena, Chile); the 2.2\,m telescope at the Calar Alto Observatory (Almeria, Spain); and the VLT at Paranal Observatory (Chile).
This work has made use of data from the European Space Agency (ESA) mission
{\it Gaia} (\url{https://www.cosmos.esa.int/gaia}), processed by the {\it Gaia}
Data Processing and Analysis Consortium (DPAC,
\url{https://www.cosmos.esa.int/web/gaia/dpac/consortium}). Funding for the DPAC
has been provided by national institutions, in particular the institutions
participating in the {\it Gaia} Multilateral Agreement. This publication makes use of data products from the Two Micron All Sky Survey, which is a joint project of the University of Massachusetts and the Infrared Processing and Analysis Center/California Institute of Technology, funded by the National Aeronautics and Space Administration and the National Science Foundation.
This project has made use of \textsc{Python}’s \textsc{lmfit} and \textsc{SciPy} libraries.
CG acknowledges support from the Agencia Estatal de Investigaci\'on del Ministerio de Ciencia e Innovaci\'on (AEI-MCINN) under grant “At the forefront of Galactic Archaeology: evolution of the luminous and dark matter components of the Milky Way and Local Group dwarf galaxies in the {\it Gaia} era” with reference PID2023-150319NB-C21/10.13039/501100011033. Co-funded by the European Union (ERC-2022-AdG, "StarDance: the non-canonical evolution of stars in clusters", Grant Agreement 101093572, PI: E. Pancino). Views and opinions expressed are however those of the author(s) only and do not necessarily reflect those of the European Union or the European Research Council. Neither the European Union nor the granting authority can be held responsible for them.



\section*{Data availability}
All data required to reproduce the results presented in this paper are publicly available, except for the line-strength measurements, which will be published shortly via the CDS. The processed spectra will be made available by the authors upon reasonable request.


\bibliographystyle{mnras}
\bibliography{example} 



\clearpage

\appendix

\section{APPENDICES}


Table~\ref{tab:ref_field_stars} includes the reference values for field stars: reddening, distance and metallicity.
\begin{table}
\setlength{\tabcolsep}{1mm}
\caption{Reference values for the field stars in our sample.}
\centering
\label{tab:ref_field_stars}
\begin{tabular}{lccccccccc}
\hline
            SimbadName     &    [Fe/H]       & Ref& E(B - V ) & (m-M)$_0$ & Ref\\  
\hline
 BPS CS 22172-0002    & $-3.86 \pm 0.02$ & 2  &    0.072 & $13.41 \pm 0.19$ &     14 \\
 BPS BS 16467-062    & $-3.77 \pm 0.06$ &  2 &    0.018 & $13.27 \pm 0.16$ &     14 \\
 HE1116-0634    & $-3.73 \pm 0.12$ &  6 &    0.055 & $13.82 \pm 0.22$ &     14 \\
 BPS BS 16550-0087    & $-3.53 \pm 0.09$ &  3 &    0.027 & $16.80 \pm 0.41$ &     14 \\
 BPS BS 16477-003    & $-3.36 \pm 0.02$ & 2  &    0.026 & $16.22 \pm 0.50$ &     14 \\
 BPS BS 16929-0005    & $-3.34 \pm 0.03$ & 3  &    0.011 & $12.72 \pm 0.10$ &     14 \\
 HE0401-0138    & $-3.34 \pm 0.13$ &  1 &    0.219 & $14.46 \pm 0.20$ &     14 \\
 BPS CS 30325-094 & $-3.3 \pm 0.02$  & 2  &    0.041 & $12.24 \pm 0.08$ &     14 \\
 BPS CS 22878-0101   & $-3.25 \pm 0.04$ & 2  &    0.098 & $15.70 \pm 0.32$ &     14 \\
 HD115444      & $-3.15 \pm 0.11$ &  4 &    0.014 & $9.70 \pm 0.03$  &     14 \\
 HE1311-0131    & $-3.15 \pm 0.12$ & 6  &    0.03 & $12.54 \pm 0.11$ &     14 \\
 BPS BS 16085-0050   & $-3.15 \pm 0.13$ &  12 &    0.024 & $13.18 \pm 0.23$ &     14 \\
 BPS CS 30312-0059    & $-3.14 \pm 0.09$ &  3 &    0.119 & $12.96 \pm 0.14$ &     14 \\
 HE1317-0407    & $-3.1 \pm 0.12$  & 6  &    0.036 & $13.84 \pm 0.27$ &     14 \\
 BPS BS 16080-054    & $-3.07 \pm 0.09$ &  3 &    0.028 & $13.65 \pm 0.12$ &     14 \\
 BPS BS 16928-0053    & $-3.07 \pm 0.09$ & 3  &    0.011 & $15.90 \pm 0.36$ &     14 \\
 BD-185550      & $-3.06 \pm 0.02$ & 2  &    0.191 & $8.57 \pm 0.02$  &     14 \\
 HE0420+0123    & $-3.03 \pm 0.13$ & 6  &    0.163 & $10.38 \pm 0.05$ &     14 \\
 HD237846      & $-3.01 \pm 0.1$  & 13  &    0.01  & $9.69 \pm 0.02$  &     14 \\
 HD88609       & $-2.97 \pm 0.13$ &  4 &    0.01  & $10.41 \pm 0.05$ &     14 \\
 HE1254+0009    & $-2.94 \pm 0.13$ &  1 &    0.021 & $16.89 \pm 0.52$ &     14 \\
 BPS CS 22877-0011   & $-2.92 \pm 0.1$  &  11 &    0.038 & $12.92 \pm 0.14$ &     14 \\
 HE1311-1412    & $-2.91 \pm 0.13$ &  1 &    0.082 & $17.71 \pm 0.58$ &     14 \\
 HE1252-0117    & $-2.89 \pm 0.13$ & 1  &    0.023 & $16.19 \pm 0.44$ &     14 \\
 HD122563      & $-2.82 \pm 0.04$ & 2  &    0.024 & $7.57 \pm 0.02$  &     14 \\
 BPS CS 22175-0007    & $-2.81 \pm 0.18$ & 1  &    0.026 & $12.93 \pm 0.11$ &     14 \\
 HE1320-1339    & $-2.78 \pm 0.13$ &  1 &    0.065 & $11.38 \pm 0.07$ &     14 \\
 BPS CS 31082-0001    & $-2.78 \pm 0.19$ &  1 &    0.015 & $11.84 \pm 0.13$ &     14 \\
 HE1225+0155    & $-2.75 \pm 0.13$ & 1  &    0.02  & $12.81 \pm 0.14$ &     14 \\
 BD+053098      & $-2.74 \pm 0.13$ & 4  &    0.068 & $9.95 \pm 0.03$  &     14 \\
 HD107752      & $-2.66 \pm 0.13$ & 13  &    0.031 & $10.79 \pm 0.06$ &     14 \\
 HD186478      & $-2.59 \pm 0.02$ & 2  &    0.126 & $9.87 \pm 0.04$  &     14 \\
 BD+233130      & $-2.58 \pm 0.03$ & 3  &    0.064 & $7.35 \pm 0.01$  &     14 \\
 HD85773       & $-2.58 \pm 0.1$  &  10 &    0.046 & $11.78 \pm 0.07$ &     14 \\
 BD+042621      & $-2.52 \pm 0.05$ & 4  &    0.02  & $10.97 \pm 0.08$ &     14 \\
 HE0243-0753    & $-2.49 \pm 0.13$ &  1 &    0.026 & $14.82 \pm 0.19$ &     14 \\
 HE0442-1234    & $-2.41 \pm 0.13$ & 1  &    0.161 & $13.94 \pm 0.21$ &     14 \\
 HD108317      & $-2.35 \pm 0.13$ & 7  &    0.018 & $6.46 \pm 0.01$  &     14 \\
 HD165195      & $-2.32 \pm 0.13$ & 10  &    0.196 & $9.22 \pm 0.03$  &     14 \\
 HD110184      & $-2.25 \pm 0.13$ & 8  &    0.022 & $11.16 \pm 0.08$ &     14 \\
 BD-012582      & $-2.25 \pm 0.1$  & 10  &    0.021 & $7.97 \pm 0.02$  &     14 \\
 BD+371458      & $-2.17 \pm 0.07$ & 7  &    0.281 & $5.81 \pm 0.00$  &     14 \\
 HD103545      & $-2.14 \pm 0.13$ &  10 &    0.037 & $9.80 \pm 0.04$  &     14 \\
 HD87140       & $-1.95 \pm 0.07$ &  7 &    0.006 & $7.59 \pm 0.01$  &     14 \\
 BD+042466      & $-1.92 \pm 0.05$ &  13 &    0.042 & $10.56 \pm 0.05$ &     14 \\
 HD63791       & $-1.72 \pm 0.13$ & 4  &    0.054 & $7.80 \pm 0.02$  &     14 \\
 HD74462       & $-1.56 \pm 0.07$ & 10  &    0.057 & $8.68 \pm 0.02$  &     14 \\
 BD+302034      & $-1.53 \pm 0.05$ & 9  &    0.023 & $12.77 \pm 0.13$ &     14 \\
 HD105546      & $-1.49 \pm 0.13$ & 10  &    0.023 & $8.32 \pm 0.02$  &     14 \\
 HD15656       & $-0.16 \pm 0.17$ &  5 &    0.102 & $5.63 \pm 0.03$  &     14 \\
\hline
\end{tabular}
References: (1) \cite{2005Barklem}; (2) \cite{2011Andrievsky}; (3) \cite{2008LaiL}; (4) \cite{2002Johnson}; (5) \cite{1990McWilliam}; (6) \cite{2011Hollek}; (7) \cite{2000Fulbright}; (8) \cite{2011Wu}; (9) \cite{1985Luck}; (10) \cite{1996Pilachowski}; (11) \cite{1995McWilliam}; (12) \cite{2001Giridhar}; (13) \cite{2009Zhang}; (14) \cite{Gaia2021}.
\end{table}


\bsp	
\label{lastpage}
\end{document}